\documentclass{ws-procs961x669}            % default, citations in superscript
\pdfoutput=1

\renewcommand\Re{\operatorname{\mathfrak{Re}}}
\renewcommand\Im{\operatorname{\mathfrak{Im}}}

\begin{document}
\title{Extraction of DVCS form factors with uncertainties}
\footnotetext{Contribution to proceedings of INT program 18-3, University of Washington, Seattle}

\author{K. Kumeri\v{c}ki}

\address{
    Department of Physics, Faculty of Science, University of Zagreb, 10000 Zagreb, Croatia\\
    Institut f\"{u}r Theoretische Physik, Universit\"{a}t Regensburg, D-93040 Regensburg, Germany
     }

\begin{abstract}
    We discuss recent attempts to extract deeply virtual Compton scattering 
    form factors with emphasis on their uncertainties, which turn out to
    be most reliably provided by method of neural networks.
\end{abstract}


\section{Introduction}

Partonic structure of the nucleon, as encoded by generalized parton distributions
(GPDs), is essentially non-perturbative. As such, main avenue
to its determination is extraction from experimental data,
%Procedure is in principle equivalent to time-tested procedure for determination of standard
%forward parton distributions (PDFs): one chooses some parametrization of
%non-perturbative function and fits it to the data. 
mostly from measurements of deeply virtual Compton scattering (DVCS),
which is a subprocess of electroproduction of real photon off nucleon.
Still, more than a decade after the first such fitting attempts\cite{Kumericki:2007sa},
we have only partial phenomenological knowledge of GPDs. (Recent review is
available in Ref.~\citenum{Kumericki:2016ehc}.)
Furthermore, although assessment of uncertainties is indispensable part
of any quantitative scientific result, authors of global GPD fits usually hesitated to 
discuss error bands of extracted functions.
It was understood that standard simple propagation of experimental uncertainties is
not enough. GPD functions depend in
a rather unknown manner on three kinematic variables (average
and transferred parton longitudinal momentum fractions, $x$ and $\xi$, 
and nucleon momentum transfer squared $t$), which makes the
problem very complex from the data-analysis standpoint, and
the very choice of fitting parametrization introduces unknown and
possibly dominant uncertainty.

\section{DVCS subtraction constant}

\begin{figure}[h]
\begin{center}
    \includegraphics[width=0.9\textwidth]{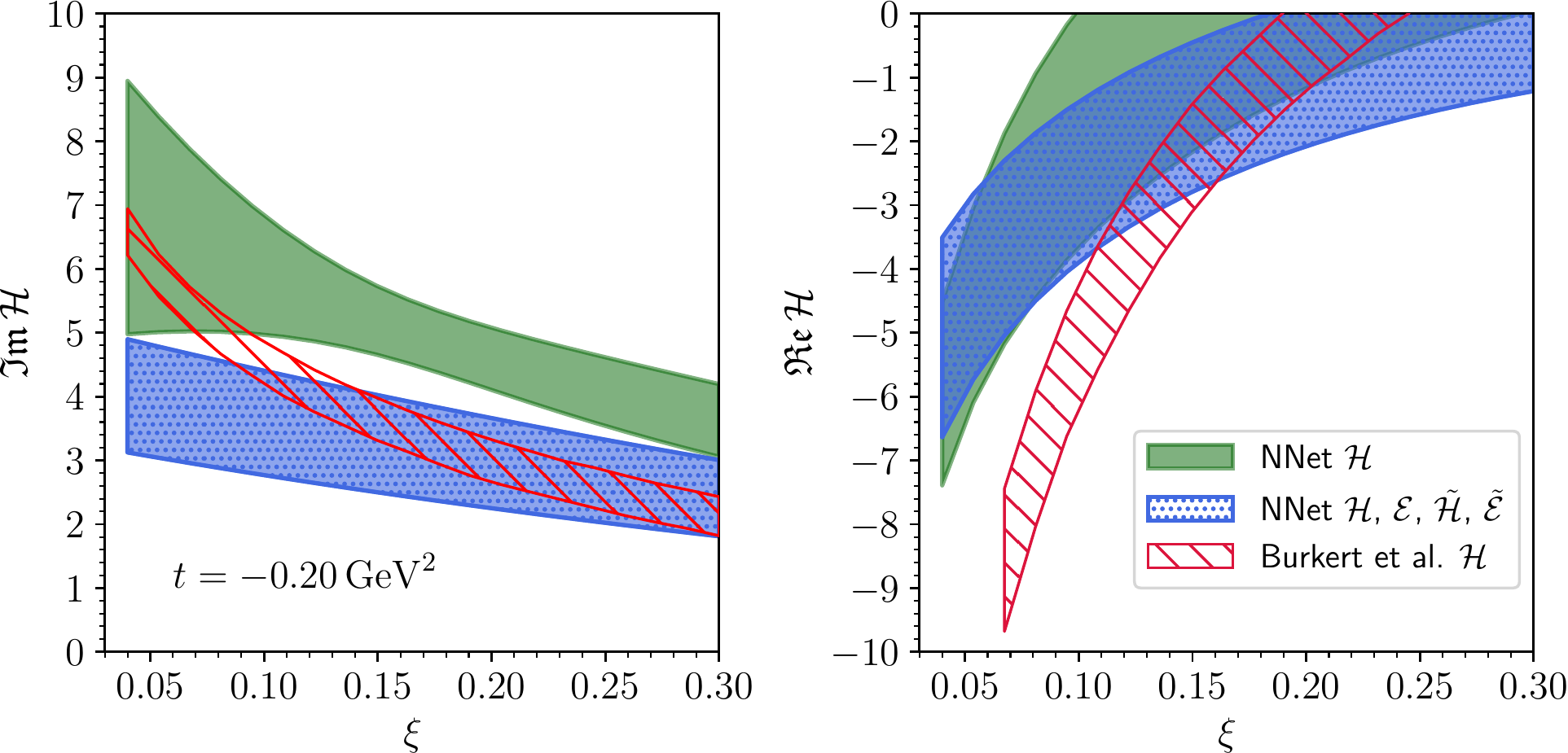}
\end{center}
\caption{Imaginary and real part of CFF $\mathcal{H}$ resulting from
fitting different parametrizations to the essentially the same data.}
\label{fig:nat}
\end{figure}

Important role of the choice of the parametrization may be illustrated by recent
attempts to determine the subtraction constant $\Delta(t)$ of DVCS dispersion
relation, 
\begin{equation}
\Re\mathcal{H}(\xi,t) = \Delta(t) + \frac{1}{\pi} {\rm P.V.}
\int_{0}^{1} {\rm d}x \left(\frac{1}{\xi-x} - \frac{1}{\xi+x}\right)\Im\mathcal{H} (x,t) \,,
\label{eq:disp}
\end{equation}
that is of great phenomenological interest since it is
closely related to the pressure in the nucleon\cite{Polyakov:2002yz,Teryaev:2005uj}.
Compton form factor (CFF) $\mathcal{H}(\xi,t)$ in Eq. (\ref{eq:disp})
is a convolution of
GPD $H(x,\xi,t)$ with the known hard scattering amplitude and is, being dependent
on two variables only, more easy extraction target.
Still, $\Delta(t)$ resulting from fits to CLAS DVCS 
data\cite{Jo:2015ema}, came out with very different uncertainty estimation,
depending on whether relatively rigid ansatz\cite{Kumericki:2009uq} for $\mathcal{H}$ was
used\cite{Burkert:2018bqq} or it was parametrized by completely
flexible neural networks\cite{Kumericki:2019ddg}, see Fig. \ref{fig:nat}.

\section{Neural network fits}
In the lack of general procedure for assessment of systematic uncertainties
coming from the choice of fitting ansatz, one convenient approach
is using the parametrization by neural networks,
which is known \emph{not} to introduce any such systematic error.
After the early proof of concept\cite{Kumericki:2011rz}, first global
neural network determination of CFFs was reported in Ref. \citenum{Moutarde:2019tqa},
demonstrating the power of this approach.

Similarly, in the framework of neural net approach, we attempted 
to address the question
of which of the four leading order CFFs, $\mathcal{H}$, $\mathcal{E}$,
$\tilde{\mathcal{H}}$, and $\tilde{\mathcal{E}}$ (or, more accurately,
eight sub-CFFs which are the real and imaginary parts of these four),
can be reliably extracted from the given data.
To this end, we used the stepwise regression method proposed in
Ref. \citenum{Kumericki:2013br}, where the number of sub-CFFs is gradually
increased and all combinations are tried, until there is no statistically
significant improvement in the description of the data.
Representative subset of global DVCS data was used, with
various beam and target, spin and charge asymmetries measured
by HERMES\cite{Airapetian:2012mq,Airapetian:2008aa,Airapetian:2010ab},
and helicity independent and dependent cross-sections measured
by Hall A and CLAS JLab collaborations\cite{Defurne:2015kxq,Jo:2015ema},
where JLab data was Fourier-transformed, so that we fitted to the
total of 128 harmonics.

\begin{figure}[h]
\begin{center}
\includegraphics[width=0.95\textwidth]{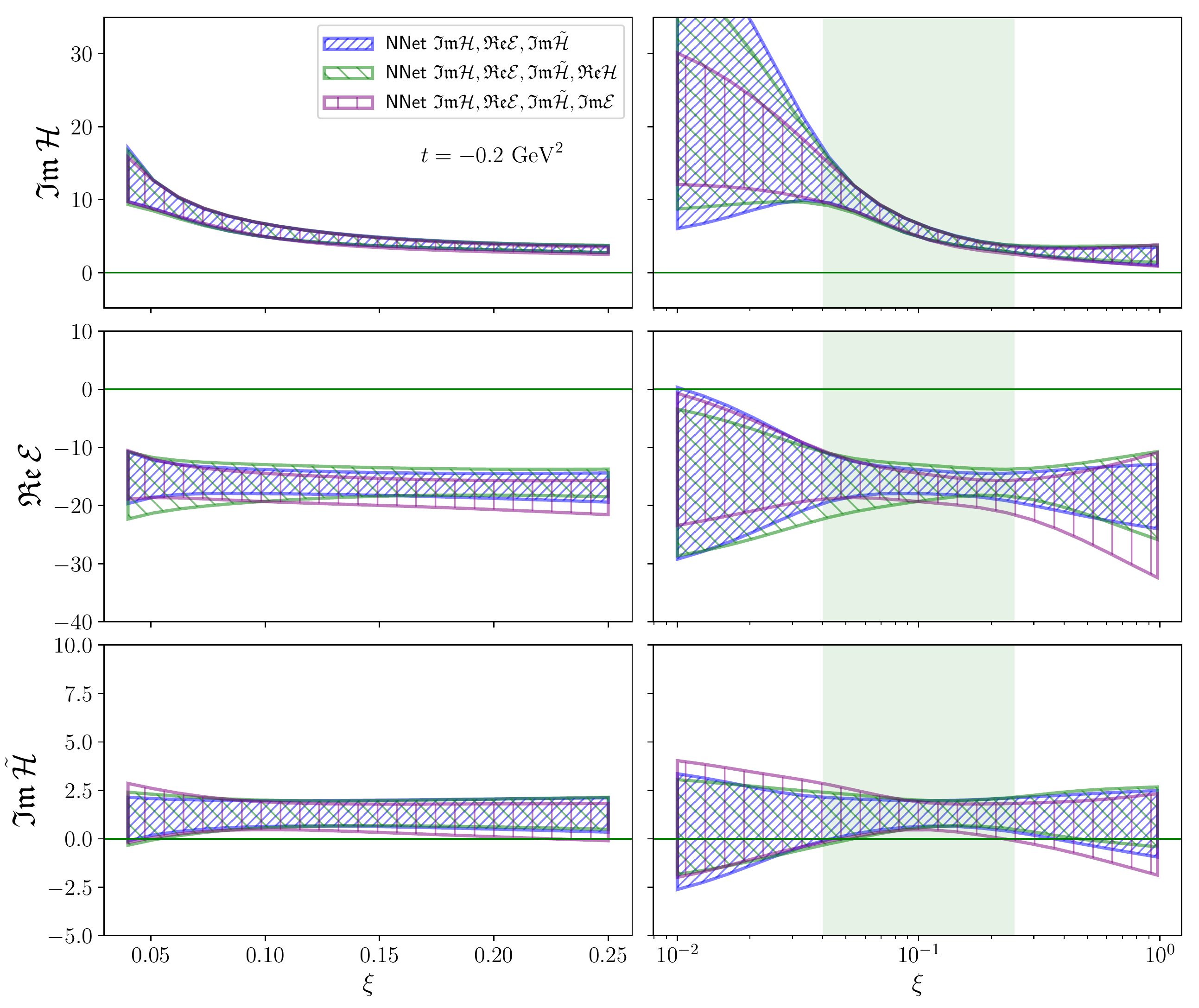}%
\end{center}
\caption{Neural network extraction of dominant CFFs from DVCS
data. Results for various sets of CFFs are consistent in the data
region (left) and also when extrapolated outside of the data region
(right). Dispersion relation constraints were \emph{not} used.}
\label{fig:left}
\end{figure}

\begin{figure}[h]
\begin{center}
\includegraphics[width=0.92\textwidth]{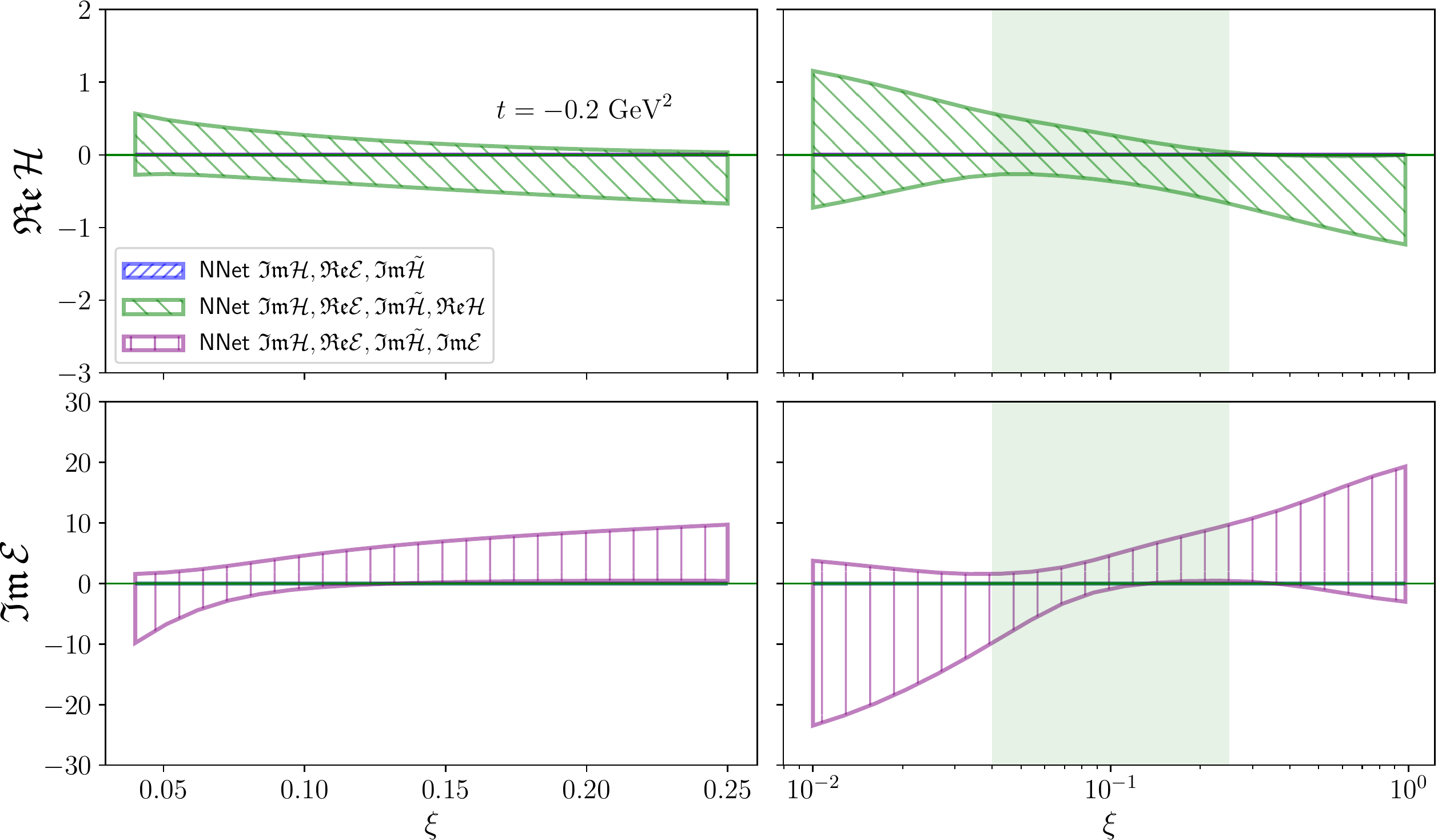}%
\end{center}
\caption{Extracted $\Re\mathcal{H}$ or $\Im\mathcal{E}$ are
mostly consistent with zero, but their addition to the model
improves description of the data from $\chi^{2}/{\rm npts}$
= 103.1/128 to 97.4/128 or 96.4/128, respectively.}
\label{fig:right}
\end{figure}

Results, displayed on Figs. \ref{fig:left} and \ref{fig:right}, show
that from the present data only $\Im\mathcal{H}$, $\Im\tilde{\mathcal{H}}$,
and $\Re\mathcal{E}$ can be reliably extracted, with maybe some
ambiguous hints of $\Re\mathcal{H}$ or $\Im\mathcal{E}$.
This is similar to the conclusions of Ref. \citenum{Kumericki:2013br}, which
used method of local fits (which is also resistant to the problem
of choice of the ansatz function).

\section{Conclusion}
How to reliably determine uncertainties of GPD or CFF functions
extracted by fitting of ansatz function is an important open
question for this area of research. At the moment, the best
confidence is provided by the method of neural networks.

\section*{Acknowledgments}

This work was supported by 
QuantiXLie Centre of Excellence through grant KK.01.1.1.01.0004,
and European Union's Horizon 2020
research and innovation programme under grant agreement No 824093.

%\section{References}

%\bibliographystyle{ws-procs961x669}
%\bibliography{\BibPath/kkumer-ascii}

%Non BiBTeX users can list down their references as:

\providecommand{\href}[2]{#2}\begingroup\raggedright\endgroup

\end{document}